\documentclass[12pt,openbib]{article}
\usepackage{a4}
\usepackage{amssymb,theorem}
\theorembodyfont{\normalfont}

\newcommand{\R}{\mathbb{R}}

\newcommand{\SO}{\mathop{\mathrm{SO}}\nolimits}
\newcommand{\oo}{\mathop{\mathrm{\! \, o}}\nolimits}
\newcommand{\Oo}{\mathop{\mathrm{\! \, O}}\nolimits}

\newcommand{\Ad }{\mathop{\mathrm{Ad}}\nolimits}

\newcommand{\comp}{\raisebox{0pt}{$\scriptstyle\circ \, $}}
\newcommand{\setrule}{\, \rule[-4pt]{.5pt}{13pt}\, }

\newcommand{\smallrowspace}{\rule{0pt}{14pt}}
\newcommand{\rowspace}{\rule{0pt}{16pt}}

\newcommand{\tfrac}[2]{\mbox{${\scriptstyle \frac{{#1}}{{#2}}}$}}
\newcommand{\onehalf}{\mbox{$\frac{\scriptstyle 1}{\scriptstyle 2}\,$}}
\newcommand{\spann}{\mathop{\rm span}\nolimits}

\begin{document}
\mbox{}
\vspace{.4in}
\begin{center}
{\Large \bf A new interpretation for the mass of a\\
\rule{0pt}{18pt}  classical relativistic particle } \\
\mbox{} \vspace{.05in} \\
\parbox[t]{3.75in}{Richard Cushman\footnotemark \hspace{2pt}and
Wilberd van der Kallen\footnotemark \\
\mbox{}\vspace{-.1in} \\
Mathematics Institute, University of Utrecht, 3508TA Utrecht,
The Netherlands}
\end{center}
\addtocounter{footnote}{-1}
\footnotetext{email: cushman@math.uu.nl}
\addtocounter{footnote}{1}
\footnotetext{email: vdkallen@math.uu.nl}
\begin{abstract}
Based on a recent classification
of coadjoint orbits of the full Poincar\'{e} group, we give a new group
theoretic interpretation for the mass of a classical relativistic
particle.
\end{abstract}
\addtocounter{footnote}{1}
\footnotetext{\parbox[t]{4in}{Mathematics Subject Classifications (2000) 20E45,
20G45, 22E15. }}

In the late nineteen sixties
there was some work on the classification of coadjoint orbits of the
Poincar\'{e} group done in \cite{renouard} using Wigner's little
group method. The thesis concentrated on
geometric quantization, as did Rawnsley \cite{rawnsley}, who considered
a general semidirect product of a Lie group and a vector space. \medskip

Using the recent classification of all the coadjoint orbits
of the Poincar\'{e} group in \cite{cushman-vanderkallen},
we obtain a new interpretation for the mass of
a classical relativistic particle \cite{souriau},
\cite{arens}, and \cite{guillemin-sternberg}. \medskip

We interpret nonzero mass of a classical relativistic particle to
be a nonzero \emph{modulus} for the coadjoint orbit
of the full Poincar\'{e} group which represents the particle.
To explain what we mean by a modulus, consider the
group $G = \{ ${\footnotesize $\left(
\begin{array}{cc} 1 & a \\ 0 &1 \end{array} \right) \setrule $
$ a \in \R $}$\} $.
Since $G$ is abelian, {\footnotesize $\left(
\begin{array}{cc} 1 & a \\ 0 &1 \end{array} \right) $} is
conjugate to {\footnotesize $\left(
\begin{array}{cc} 1 & a' \\ 0 &1 \end{array} \right) $}
in $G$ if and only if $a=a'$. Thus the conjugacy classes of elements
of $G$ are parametrized by the real number $a$, which is clearly not an
eigenvalue of
any element of $G$. For the full Poincar\'{e} group
our modulus is
independent of the component of the coadjoint orbit. In our classification it
is consistent to take it positive, and we did so.
Since mass has been measured to be positive,
we interpret mass as being the modulus, not its negative.
It follows that in our interpretation there
are \emph{no} classical relativistic particles with \emph{negative} mass.
Thus we partly resolve a difficulty which plagued Souriau.
(See the last footnote on page 192  of \cite{souriau}.)
The Lorentz length squared of the energy-momentum vector of the
particle is still the square of the mass. \medskip

The components of the coadjoint orbit of a classical relativistic
particle are labeled by physical
data which characterize the particle; namely,
the sign of the spin for a massive classical relativistic particle
with spin; the sign of the energy for a massive classical relativistic particle
without spin; the sign of the energy and the helicity for a massless
classical relativistic particle. \medskip

Let us recall what mass means in the standard classification of
classical relativistic particles. In \cite{barut-raczka} an elementary system
=(classical relativistic particle) is the representation space (which
is a specific Hilbert space) of a single irreducible unitary
representation of the full Poincar\'{e} group. Such a representation
is characterized by mass squared $(m^2)$ and spin $(j(j+1))$ or helicity.
There is a list in \cite{barut-raczka} of irreducible unitary
representations of the universal covering group of the identity component of
the Poincar\'{e} group. A complete list of the irreducible unitary
representations of the \emph{full Poincar\'{e} group} do not seem to be known,
in spite of the ``folklore'' in the physics community that they are,
\cite{bargmann-wigner},
\cite{wigner}. In \cite{souriau} an elementary system is a connected
component of a coadjoint orbit of the restricted Poincar\'{e} group
(= the identity component of the full Poincar\'{e} group).
The coadjoint orbits of the restricted Poincar\'{e} group are classified
by mass (= the signed square roots of the Lorentz length of the energy
momentum vector), spin or
helicity, \cite{souriau}. The relation
between the representation theoretic and coadjoint orbit definitions of
an elementary system is
the procedure of geometric quantization,
see \cite{souriau,rawnsley}, which constructs an irreducible unitary
representation starting from a coadjoint orbit. \bigskip

\noindent {\large \bf \S 1. Facts about the Poincar\'{e} group}
\bigskip

First we recall some standard facts about the Poincar\'{e} group, which
we will use later on. Let $\{ e_1, e_2,e_3, e_4 \}$ be a basis of
Minkowski space $({\R }^4, \gamma )$ such that the Gram matrix
of the Lorentz inner product $\gamma $ is $G =\mathrm{diag}\, (-I_3, 1)$,
that is, $\gamma $ is diagonal with signature $-\, - \, - \, +$. The
Lorentz group $\Oo (3,1)$ is the Lie group of linear maps of
${\R }^4$ into itself which preserve $\gamma $. The Lie algebra
$\oo (3,1)$ of $\Oo (3,1)$ consists of $4 \times 4$ real
matrices $M$ such that $M^TG +GM =0$, that is,
$M=${\tiny $\left( \begin{array}{cc} \widehat{\ell } & g \\
g^T & 0 \end{array} \right) $}, where $\ell = ({\ell }_1, {\ell }_2,
{\ell }_3), \, g \in {\R }^3 = \spann \{ e_1, e_2, e_3 \}$ and
$\widehat{\ell }=${\tiny  $\left( \begin{array}{rrr} 0 & - {\ell }_3 &
{\ell }_2 \\ {\ell }_3 & 0 & -{\ell }_1 \\ -{\ell }_2 & {\ell }_1 & 0
\end{array} \right) $.} The Poincar\'{e} group $\mathcal{P}$ is the
group of \emph{affine} linear Lorentz transformations of Minkowski space
into itself, that is, $\mathcal{P}$ is the semidirect product of the
Lorentz group and the abelian group ${\R }^4$. We can write
$(S,C) \in \mathcal{P} \subseteq \Oo (3,1) \times {\R }^4$ as the $5 \times 5$
real matrix {\tiny $\left( \begin{array}{cc} S & C \\ 0 & 1 \end{array}
\right) $.} Group multiplication $\cdot$ in $\mathcal{P}$ is then given
by matrix multiplication.  An element $(\Sigma , \Gamma )$ of the Lie algebra
$\wp \subseteq \oo (3,1) \times {\R }^4$ of the Poincar\'{e}
group can be written as the $5 \times 5$
real matrix {\tiny $\left( \begin{array}{cc} \Sigma  & \Gamma \\ 0 & 0
\end{array} \right) $.} \medskip

Using the nondegenerate inner product
\begin{equation}
\langle \, \, | \, \, \rangle : \wp \times \wp \rightarrow \R :
((M,P), (\Sigma , \Gamma )) \mapsto -\onehalf \, \mathrm{tr}\,
M\Sigma - \gamma (P, \Gamma ),
\label{eq-s2one}
\end{equation}
we may identify $\wp $ with its dual space ${\wp }^{\ast }$.
Thus we think of $(M,P)$ as an element $\nu $ of ${\wp }^{\ast }$. Under this
identification, the coadjoint
action
\begin{displaymath}
\mathcal{P} \times {\wp }^{\ast } \rightarrow {\wp }^{\ast }:
((S,C), \nu ) \mapsto {\Ad }^T_{(S,C)^{-1}}\nu
\end{displaymath}
of $\mathcal{P}$ on ${\wp }^{\ast }$ becomes the action
\begin{equation}
\widetilde{\Ad }: \mathcal{P} \times \wp  \rightarrow \wp :
((S,C), (M,P) ) \mapsto  (SMS^{-1} +L_{C,SP}, SP).
\label{eq-s2two}
\end{equation}
$L_{C, SP}$ is the linear map of ${\R }^4$ into itself defined by
\begin{displaymath}
L_{C,SP}\Gamma = \gamma (SP, \Gamma )C - \gamma (C,\Gamma )SP.
\end{displaymath}
Note that $L_{C,SP} \in \oo (3,1)$. The energy-momentum vector
associated to $(M,P)$ is
$P=${\tiny $\left( \begin{array}{c} p \\ E \end{array} \right) $} and its
polarization vector is $W =${\tiny $\left( \begin{array}{c}
p \times g +E \, \ell \\ \langle p , \ell \rangle \end{array}
\right) $}. Here $\langle \, \, , \, \, \rangle $ is the standard Euclidean
inner product on ${\R }^3$. It is known that the Casimir
functions $C_1:\wp \rightarrow \R : (M,P) \mapsto \gamma (P,P)$ and
$C_2:\wp \rightarrow \R :(M,P) \mapsto \gamma (W,W)$ are invariant under
the action $\widetilde{\Ad }$. \medskip

Minkowski space $({\R }^4, \gamma )$ has two standard involutions:
space inversion \linebreak $I_s: {\R }^4 \rightarrow
{\R }^4:${\tiny $\left( \begin{array}{r}
r \\ t \end{array} \right) $}$\mapsto ${\tiny $\left( \begin{array}{r}
-r \\ t \end{array} \right) $}
and time reversal
$I_t: {\R }^4 \rightarrow {\R }^4:${\tiny $\left( \begin{array}{c}
r \\ t \end{array} \right) $}$\mapsto ${\tiny $\left( \begin{array}{r}
r \\ -t \end{array} \right) $}. These involutions are Lorentz transformations
that do not lie in the identity component of the Lorentz group. Therefore
$(I_s,0)$ and $(I_t,0)$ are elements of the Poincar\'{e} group which
do not lie in its identity component ${\mathcal{P}}^{\comp }$.
In fact, the group of connected components
$\mathcal{P}/{\mathcal{P}}^{\comp }$ of $\mathcal{P}$ is
$\{ e, \, I_s, \, I_t, \, I_s\mbox{\raisebox{1pt}{$\comp $}}I_t \} $. A
calculation shows that
\begin{displaymath}
(M_s,P_s) = {\widetilde{\Ad }}_{(I_s,0)}(M,P) =
\left( \rowspace \right. \mbox{{\footnotesize $ \left(
\begin{array}{cr} \widehat{\ell } & -g \\ -g^T & 0 \end{array}
\right) $}},  \mbox{{\footnotesize $ \left(
\begin{array}{r} -p \\ E \end{array} \right) $}} \left. \rowspace \right)
\end{displaymath}
and
\begin{displaymath}
(M_t,P_t) = {\widetilde{\Ad }}_{(I_t,0)}(M,P) =
\left( \rowspace \right. \mbox{{\footnotesize $ \left(
\begin{array}{cr} \widehat{\ell } & -g \\ -g^T & 0 \end{array}
\right) $}},  \mbox{{\footnotesize $ \left(
\begin{array}{r} p \\ -E \end{array} \right) $}} \left. \rowspace \right) .
\end{displaymath}
Therefore the polarization vector $W_s$ corresponding to
$(M_s,P_s)$ is {\tiny $ \left( \begin{array}{c} p \times g + E\, \ell  \\ -
\langle \ell , p \rangle \end{array} \right) $}; whereas the polarization
vector $W_t$ corresponding to $(M_t,P_t)$ is {\tiny  $ \left(
\begin{array}{c} -p \times g - E\, \ell  \\ \langle \ell , p \rangle
\end{array} \right) $.}
\bigskip

\noindent {\large \bf \S 2. The classification} \bigskip

 From the classification of the coadjoint orbits of the Poincar\'{e}
group given in \cite{cushman-vanderkallen}, we extract
the following physically significant coadjoint orbits, following
the criteria in \cite{souriau}. \medskip

\noindent 1.  Let $\mathcal{O}$ be the coadjoint orbit containing
the covector $\nu = (M^{\comp }, P^{\comp }) =$ \linebreak
{\tiny $ \left( \smallrowspace
\right. \left( \begin{array}{cc} \beta \, {\widehat{e}}_3 & 0 \\
0 & 0 \end{array} \right) $}, $\mu \, e_4 \left. \smallrowspace \right) $,
where $\mu >0$ is a modulus. The polarization
vector associated to $\nu $ is $W^{\comp } = \mu \beta \, e_3$.
The value of the Casimir $C_1$ at $\nu $ is
${\mu }^2$ and the value of $C_2$ at $\nu $ is $-{\mu }^2{\beta }^2$.
Therefore the classical
relativistic particle corresponding to $\mathcal{O}$ has positive
mass $\mu $ and spin $\pm \beta = \pm \sqrt{\frac{-\gamma
(W^{\comp }, W^{\comp })}{\gamma (P^{\comp }, P^{\comp })}}$. At $\nu $ the
isotropy group
\begin{displaymath}
{\mathcal{P}}_{(M^{\comp }, P^{\comp })} =
\{ (S,C) \in \mathcal{P} \setrule \,
{\widetilde{\Ad }}_{(S,C)}(M^{\comp },P^{\comp }) =
(M^{\comp },P^{\comp }) \} ,
\end{displaymath}
given by
\begin{displaymath}
\{  (S,C), \, \,
(I_s,0)\cdot (S,C) \in \mathcal{P} \setrule
(S,C)= \mbox{{\tiny $ \left( \smallrowspace
\right. \left( \begin{array}{crcc} \cos \vartheta  &  -\sin \vartheta & 0 & 0
\\
\sin \vartheta & \cos \vartheta & 0 & 0 \\
0 & 0 & 1 & 0 \\ 0 & 0 & 0 & 1 \end{array} \right) $}}, \, \lambda \, e_4
\left.
\smallrowspace \right)
\end{displaymath}
with $\vartheta , \, \lambda \in \R $, has two connected components. Thus the
coadjoint orbit $\mathcal{O}$ has two connected components
${\mathcal{O}}^{\pm }$, which
are interchanged under time reversal. In other words, if we
assume that $(M^{\comp}, P^{\comp }) \in {\mathcal{O}}^{+}$, then
$(M^{\comp}_t, P^{\comp }_t) \in {\mathcal{O}}^{-}$ and conversely. Assuming
that the value of spin is $\beta $ on ${\mathcal{O}}^{+}$,
the value of spin on ${\mathcal{O}}^{-}$ is $- \beta $, since
\begin{displaymath}
W^{\comp }_t =  \, - W^{\comp } \, = \,
\mu (-\beta )e_3.
\end{displaymath}
\noindent 2. Let $\mathcal{O}$ be the coadjoint orbit containing
$\nu = (M^{\comp }, P^{\comp }) = (0, \mu \, e_4)$, where $\mu >0$
is a modulus. The
polarization vector corresponding to $\nu $ is $W^{\comp }=0$.
The value of the Casimir $C_1$ at $\nu $ is ${\mu }^2$ and the value
of $C_2$ at $\nu $ is $0$. Therefore the classical relativistic particle
corresponding to $\mathcal{O}$ has positive mass $\mu $ and no spin.
At $\nu $ the
isotropy group
\begin{displaymath}
{\mathcal{P}}_{(M^{\comp }, P^{\comp })} = \{  (S,C), \, \,
(I_s,0) \cdot (S,C) \in \mathcal{P} \setrule
\, (S,C)= \mbox{{\tiny $ \left( \smallrowspace
\right. \left( \begin{array}{cc} A  &  0 \\
0 & 1 \end{array} \right) $}}, \lambda \, e_4 \left. \smallrowspace \right) \}
\end{displaymath}
with $A \in \SO (3, \R )$ and $\lambda \in \R $ has two connected components.
Therefore the coadjoint orbit $\mathcal{O}$ has two connected components
${\mathcal{O}}^{\pm }$, which are interchanged by time reversal $I_t$.
Assuming that $(M^{\comp }, P^{\comp }) \in {\mathcal{O}}^{+}$, then
$(M^{\comp }_t, P^{\comp }_t) \in
{\mathcal{O}}^{-}$ and conversely. Assuming that energy $E$ is positive on
${\mathcal{O}}^{+}$, then it is negative on ${\mathcal{O}}^{-}$, because
$P^{\comp }_t = -P^{\comp } \, = \, -\mu \, e_4$. Consequently, the classical
relativistic particle corresponding
to the coadjoint orbit $\mathcal{O}$ has positive mass $\mu $, no spin,
and positive or negative energy. \medskip

\noindent 3. Let $\mathcal{O}$ be the coadjoint orbit containing
$\nu = (M^{\comp }, P^{\comp }) =${\tiny $ \left( \smallrowspace
\right. \left( \begin{array}{cc} \beta \, {\widehat{e}}_1 & 0 \\
0 & 0 \end{array} \right) $}, \linebreak
 $\tfrac{1}{\sqrt{2}}(e_1+ e_4) \left.
\smallrowspace \right) $, where $\beta >0$. The polarization vector
corresponding to
$\nu $ is $W^{\comp } = \tfrac{\beta }{\sqrt{2}}(e_1+e_4)
= \beta \, P^{\comp }$. The value of the Casimir $C_1$ at $\nu $
is \linebreak
$\gamma (\tfrac{1}{\sqrt{2}}(e_1+ e_4),\tfrac{1}{\sqrt{2}}(e_1+ e_4)) =
\tfrac{1}{2} - \tfrac{1}{2} =0$ and the value of the Casimir
$C_2$ at $\nu $ is $0$. Thus the classical relativistic particle corresponding
to $\mathcal{O}$ has no mass, but its spin is $\beta $.
A calculation shows that ${\mathcal{O}}$ has
four connected components ${\mathcal{O}}^{\pm ,\pm }$. Space inversion
$I_s$ interchanges ${\mathcal{O}}^{+,+}$ and ${\mathcal{O}}^{-,+}$.
Time inversion $I_t$ interchanges ${\mathcal{O}}^{+,+}$ and
${\mathcal{O}}^{+,-}$. Thus space time inversion
$I_s$\raisebox{1pt}{$\comp$}$I_t$ interchanges
${\mathcal{O}}^{+,+}$ and ${\mathcal{O}}^{-,-}$. Assuming that
$(M^{\comp }, P^{\comp }) \in {\mathcal{O}}^{+,+}$, then time reversal $I_t$
changes the sign of the energy, since $P^{\comp }_t =
\tfrac{1}{\sqrt{2}}(e_1- e_4) $. Thus the second $\pm $ sign in
${\mathcal{O}}^{+,+}$ is the sign of the energy.
Since $W^{\comp }_{s} =
\tfrac{\beta }{\sqrt{2}}(e_1-e_4)=-\beta P^{\comp }_s$, space reversal
changes the direction
of the polarization vector relative to the energy-momentum vector. Thus
the first sign in ${\mathcal{O}}^{+,+}$ is the helicity. Consequently,
$\mathcal{O}$ corresponds to a classical relativistic particle with no mass,
but with positive spin, positive or negative energy, and positive or negative
helicity. \bigskip

\noindent {\large \bf \S 3. Normal form}\bigskip

Since our classification of classical nonrelativistic particles depends on
the existence of a new invariant of the coadjoint action
of the Poincar\'{e} group, it behooves us to give a method for
calculating it. We use ideas from \cite{cushman-vanderkallen} but give a
slightly different algorithm, which uses little cotypes rather than
affine cotypes, to determine normal forms for the
coadjoint orbits listed in \S 2. \medskip

Suppose that we are given $(M,P) \in \wp \subseteq \oo (3,1) \times {\R }^4$.
To find the normal form, we perform the following steps. \medskip

\noindent 1. Check if $P \ne 0$ and $\gamma (P,P) \ne 0$. Set
$\gamma (P,P) = \varepsilon \, {\mu }^2$ with ${\varepsilon }^2 =1$ and
$\mu >0$. Then $\mu $ is a modulus. Let $W =
{\spann \{ P \} }^{\gamma }$ be the
$\gamma $-orthogonal complement in ${\R }^4$ of the subspace spanned by
$P$. There are two cases.
\par
a. Check if $\varepsilon =1$ and the Gram matrix of
$\gamma |W$ is $-I_3$. Check if the characteristic polynomial of $Y = M|W$
is $\lambda ({\lambda }^2+{\beta }^2)$ for some $\beta >0$. Let $P_0$ be the
normalized eigenvector of $Y$ corresponding to the eigenvalue $0$.
Let $P_1$ be a normalized vector such that on $\spann \{ P_1,
{\beta }^{-1}\, YP_1 \} $ we have $Y^2 +{\beta }^2 =0$.
Set $P_3 = {\mu }^{-1}\, P$. Then $\{
P_1, \, {\beta }^{-1}\, YP_1, \,P_0,\, P_3 \} $ is a $\gamma $-orthogonal
basis of ${\R }^4$ with respect to which the Gram matrix of
$\gamma $ is $\mathrm{diag}\, (-I_3, 1)$ and the pair
$(M, P)$ has the normal form of case 1 of \S 2.
\par
b. Check if the characteristic polynomial of $Y = M|W$ is
${\lambda }^3$ and $Y$ is diagonalizable. Let $P_0$, $P_1$, and $P_2$
be an orthonormal system of eigenvectors of $Y$. Set
$P_3 = {\mu }^{-1}\, P$. Then $\{ P_0,\,
P_1, \, P_2, \, P_3 \} $ is a $\gamma $-orthogonal
basis of ${\R }^4$ with respect to which the Gram matrix of
$\gamma $ is $\mathrm{diag}\, (-I_3, 1)$ and the pair
$(M, P)$ has the normal form of case 2 of \S 2. \medskip

\noindent 2. Check if  $P \ne 0$ and $\gamma (P, P) = 0$.
Find a nonzero vector $\widehat{P}$ such that
$\gamma (P, \widehat{P}) =1$ and $\gamma (\widehat{P}, \widehat{P}) =0$. Let
$W = \spann \{ \widehat{P}, P \}$. The Gram matrix of $\gamma |W $
is {\tiny $ \left( \begin{array}{cc} 0 & 1 \\ 1 & 0 \end{array} \right) $.}
On $W^{\gamma }$, there is a basis $\{ Q, Q' \} $ such that the
Gram matrix of $\gamma |W^{\gamma }$ is $-I_2$. With respect to the basis
$\{ f_1, \, f_2,\, f_3, \, f_4 \} = \{ \widehat{P}, \, Q, \, Q', \, P \} $
the Gram matrix of $\gamma $ is $G'=${\tiny $\left( \begin{array}{crrc}
0 & 0 & 0 & 1 \\ 0 & -1 & 0 & 0 \\ 0 & 0 & -1 & 0 \\ 1 & 0 & 0 & 0
\end{array} \right) $} and the pair $(M,P)$ has the
form
\begin{equation}
(M',P')  \, = \, \mbox{{\footnotesize $ \left( \smallrowspace
\right. \left( \begin{array}{ccr} a & -x^TG' & 0  \\
y & Y' & x \\ 0 & -y^TG' & -a  \end{array} \right) $}},
\mbox{{\footnotesize $\left( \begin{array}{c} 0 \\ 0 \\ 1 \end{array}
\right) \left. \smallrowspace \right) . $}}
\label{eq-s4one}
\end{equation}
Here $x,y\in \spann \{ f_2, f_3 \}$, $a \in \R $, and
$(Y')^T +Y'=0$. Change coordinates using $(I_4,${\tiny $\left( \begin{array}{c}
-a \\ - y \\ 0 \end{array} \right) $}$) \in \mathcal{P}$.
 From (\ref{eq-s2two}) we see that $(M', P')$ becomes
\begin{equation}
(\widetilde{M}, \widetilde{P})  \, = \,
\mbox{{\footnotesize $ \left( \smallrowspace
\right. \left( \begin{array}{ccr} 0 & -x^TG' & 0  \\
0 & Y' & x \\ 0 & 0 & 0  \end{array} \right) $}},
\mbox{{\footnotesize $\left( \begin{array}{c} 0 \\ 0 \\ 1 \end{array}
\right) \left. \smallrowspace \right) . $}}
\label{eq-s4two}
\end{equation}\medskip

\noindent Look at the pair $(Y'= \widetilde{M}|W^{\gamma }, x)$,
where $x \in W^{\gamma }$. Check if $x=0$ and $Y' \ne 0$. Then its
characteristic polynomial on $W^{\gamma }$ is ${\lambda }^2 + {\beta }^2$ for
some $\beta >0$. There is no modulus. With
respect to the basis
\begin{displaymath}
\{ \tfrac{1}{\sqrt{2}}(f_1-f_4), \,
f_2, \, {\beta }^{-1}\, Y' f_2, \, \tfrac{1}{\sqrt{2}}(f_1+f_4) \}
\end{displaymath}
the Gram matrix of $\gamma $ is $\mathrm{diag}\, (-I_3, 1)$ and the matrix of
the pair $(\widetilde{M}, \widetilde{P})$ has the normal form of case 3
of \S 2. \medskip

To relate the coadjoint orbits in \S 2 to the classification in
\cite{cushman-vanderkallen}, we note that cases 1, 2, and 3
are denoted there by
${\nabla }^{+}_3(0),\mu + {\Delta }^{-}_0(i\beta , IP) +{\Delta }^{-}_0(0)$,
${\nabla }^{+}_3(0),\mu +{\Delta }^{-}_0(0) +{\Delta }^{-}_0(0)
+{\Delta }^{-}_0(0)$, and ${\nabla }_4(0,0) +{\Delta }^{-}_0(i\beta , IP)$,
respectively.


\begin{thebibliography}{99}

\bibitem{arens} Arens, R., Classical Lorentz invariant particles,
\textit{J. Math. Phys.}
\textbf{12} (1971) 2415--2422.

\bibitem{bargmann-wigner} V. Bargmann and E.P. Wigner, Group theoretical
discussion of relativistic wave equations,
\textit{Proc. Natl. Acad. Sci. (USA)} \textbf{34} (1948) 211--23.

\bibitem{barut-raczka} Barut, A.O. and Raczka, R.,
``Theory of Group Representations
and Applications'', 2nd edition,
World Scientific, Singapore, 1987, p.521 and p.524.

\bibitem{cushman-vanderkallen} R. Cushman and W. van der Kallen,
Adjoint and coadjoint orbits of the Poincar\'{e} group, {\tt
arXiv:math.RT/0305442}.

\bibitem{guillemin-sternberg} Guillemin, V. and Sternberg, S.,
``Symplectic Techniques in Physics'', Cambridge University Press,
Cambridge, 1984.

\bibitem{rawnsley} Rawnsley, J.H., Representations of a
semi-direct product by quantization, \textit{Math. Proc. Camb. Phil. Soc.},
\textbf{78} (1975) 345--350.

\bibitem{renouard} Renouard, P., ``Vari\'{e}t\'{e}s Symplectiques
et Quantification'', Ph.D Thesis, Universit\'{e} de Paris XI Orsay, 1969.

\bibitem{souriau} Souriau, J-M., ``Structure of Dynamical Systems'',
Birkh\"{a}user, Basel, 1997, p.174 and Chapter V.

\bibitem{wigner} E.P. Wigner, On unitary representations of the
inhomogeneous Lorentz group, \textit{Ann. of Math.} \textbf{40} (1939)
149--204.


\end{thebibliography}
\end{document}